\documentclass[aps,twocolumn,pra,showpacs,floatfix]{revtex4}
\usepackage{amsmath}
\usepackage{epsfig}
\usepackage{array}
\usepackage{natbib}

\begin{document}

\title{Magic frequencies for cesium primary frequency standard}

\author{V. V. Flambaum and  V. A. Dzuba}
\affiliation{School of Physics, University of New South Wales,
 Sydney 2052, Australia}
\author{A. Derevianko}
\affiliation {
Department of Physics, University of Nevada, Reno, Nevada 89557, USA}

\begin{abstract}
We consider microwave hyperfine transitions in the ground state of
cesium and rubidium atoms which are presently used as the primary and the secondary frequency standards. 
The atoms are confined in an optical lattice generated by a circularly
polarized laser field. We demonstrate that applying an external magnetic
field with appropriately chosen direction may cancel
dynamic Stark frequency shift making the frequency of the clock transition
insensitive to the strengths of both the laser and the magnetic fields.
This can be attained for practically any laser frequency which is sufficiently
distant from a resonance. 
%We present results for the primary Cs and secondary Rb frequency standards.
%This would also work for rubidium and other
%microwave frequency standards.
\end{abstract}
% 06.30.Ft Time and frequency
% 37.10.Jk Atoms in optical lattices
% 31.15.A- <i>Ab initio</i> calculations
\maketitle

%\section{Introduction}
The microwave transition between Cs hyperfine levels is used to define the second, the unit of time. 
The best accuracy of the Cs standard is now achieved in a fountain clock\cite{BizLauAbg05etal,HeaJefDon05etal}.
For optical clocks with neutral atoms the best results are obtained
using optical lattices where atoms are trapped in minima
(or maxima) of a laser standing wave
\cite{KatTakPal03,YeKimKat08,TakHonHig05,LeTBaiFou06,LudZelCam08etal}.
Here a laser producing an optical lattice
operates at a ``magic frequency'' where the light shift of the clock transition
is zero (due to cancellation of the light shifts of the lower and upper
clock levels).
%Magic frequencies are  known for several  optical transitions.
%(see, e.g. a review~\cite{who?}).
To extend this successful  technique to microwave
frequencies one must firstly check if the magic frequencies exist
for such transitions. In our recent work \cite{BelDerDzu08} we found the
magic frequencies for aluminium and gallium hyperfine transitions where
the valence electron is in the $p_{1/2}$ state
 (the magic frequency there is a result of mixing of the $p_{1/2}$ and
 $p_{3/2}$ states by the hyperfine interaction - see details in
 \cite{BelDerDzu08}). However, we have not found
any magic frequencies for Cs, Rb and other atoms with the $s_{1/2}$ electron
in the ground state. In Ref.~ \cite{BelDerDzu08}, we considered the case
of the linear polarization of the lattice laser light. In the present
work we show that the cancellation of the light shifts for Cs, Rb and
 other atoms with $s_{1/2}$ (or  $p_{1/2}$) electron exist if\\
1. the laser light in optical lattice has a circular (or elliptical)
 polarization,\\
2. The angle $\theta$ between the quantizing magnetic field  $B$ and  the light wave vector $k$
is close to $90^o$.\\
By varying $\theta$ one can achieve the
cancellation of the light shifts for any frequency $\omega$, i.e., any frequency
can be ``magic''. Numerical calculation of $\theta(\omega)$ has been
performed for the cesium and rubidium microwave frequency standards.

%{\em Hyperfine clock transitions in magnetic field ---}
\paragraph{Hyperfine clock transitions in magnetic field.}
Let us start from the case of no magnetic field.
The light shift of an individual atomic energy level
% of a hyperfine structure multiplet
in a circularly-polarized laser field is given by (see, e.g.,~\cite{ManOvsRap86})
\begin{eqnarray}
&&\delta E_{nFM_{F}}^{\mathrm{circ}}=-\left(  \frac{1}{2}\mathcal{E}\right)
^{2}\left[  \alpha_{nF}^{s}\left(  \omega\right)  +A~\alpha_{nF}^{a}\left(
\omega\right)  \frac{M_{F}}{2F} \right. \nonumber \\
&&\left. -\alpha_{nF}^{T}\left(  \omega\right)
\frac{3M_{F}^{2}-F\left(  F+1\right)  }{2F\left(  2F-1\right)  }\right].
\label{shift0}
\end{eqnarray}
Here $\mathcal{E}$ is the amplitude of the laser field,
 $\mathbf{F} = \mathbf{I} + \mathbf{J}$; $I$ is the nuclear spin; $J$ is the
total electron momentum; $M_F$ is projection of $\mathbf{F}$ on the
quantization axis; $\alpha^s$, $\alpha^a$ and $\alpha^T$ are the scalar,
vector (axial) and tensor polarizabilities of an atom.
The quantization axis here is along the propagation of the laser light
 and $A$ is the
degree of the circular polarization ($A=1$ for the right-hand and $A=-1$ for
the left-hand polarizations).

In the presence of a sufficiently large magnetic field the direction
of the quantization axis is along the magnetic field. In this case
the shift depends on the angle  $\theta$
between the magnetic field ${\bf B}$ and  the direction of light propagation
determined by the wave vector ${\bf k}$.  This gives an
 additional  factor $\cos{\theta}$ in the vector polarizability contribution
(the central term in (\ref{shift0})). The tensor  contribution
 (the last term in (\ref{shift0})) is multiplied by a factor  $\xi(\theta,\phi)$
which depends on orientations of the B-field and polarization;
for the circular polarization $\xi(\theta)=(3 \cos^2\theta-1)/2$.
\begin{eqnarray}
&&\delta E_{nFM_{F}}^{\mathrm{circ}}=-\left(  \frac{1}{2}\mathcal{E}\right)
^{2}\left[  \alpha_{nF}^{s}\left(  \omega\right)  +\cos{\theta}A~\alpha_{nF}^{a}\left(
\omega\right)  \frac{M_{F}}{2F} \right. \nonumber \\
&&\left. -\xi(\theta,\phi) \alpha_{nF}^{T}\left(  \omega\right)
\frac{3M_{F}^{2}-F\left(  F+1\right)  }{2F\left(  2F-1\right)  }\right].
\label{shift}
\end{eqnarray}
The effect of the laser field on the frequency of a microwave clock transition
can be found by using (\ref{shift}) for both levels and taking the difference.
We are interested in cases when this difference is zero, so that the
frequency is insensitive to the laser field. 

The relevant clock shift of a transition between hyperfine levels arises in the second (quadratic in laser field) and
the third order of perturbation theory (quadratic in laser field + linear in hyperfine interaction). 
In the case of Al and Ga hyperfine-structure (hfs)
transitions in the $p_{1/2}$ ground states considered in Ref.~\cite{BelDerDzu08},
the zero frequency shift is due to cancellation between scalar and tensor hfs-induced
polarizabilities. Clock states with $M_F=0$ were considered so that the vector part
didn't contribute. The tensor polarizabilities of atoms in the $p$ states are
strongly enhanced due to small values of fine structure interval between the
 $p_{1/2}$ and $p_{3/2}$ states which goes to the energy denominator of the
 tensor polarizability (produced by the mixing of $p_{1/2}$ and $p_{3/2}$
states by the hyperfine interaction). There is no such enhancement for
 cesium and other microwave
frequency standards based on atoms with the $s_{1/2}$ ground state.
As a consequence, there are no magic frequencies  if vector term
plays no role. This was discussed  in Ref.~\cite{BelDerDzu08}.

Note, however, that the contributions of the scalar and tensor polarizabilities
to the frequency shifts in the  $s_{1/2}$ and  $p_{1/2}$ hyperfine transitions
 is proportional
to the hyperfine interaction and appear in the third order of the
 perturbation theory while the vector polarizability contribution
exists even without the hyperfine interaction, so it appears in the lower
(second) order of the perturbation theory. This means that for
 $A \cos{\theta} \sim 1$  the vector contribution
is orders of magnitude larger. Therefore, by varying angle $\theta$
(or degree of the  circular polarization $A$)
 we can always find some small  value
of the factor  $A \cos{\theta}$ in the vector contribution
 to cancel the small scalar and tensor contributions and
 make the total light shift of the hyperfine frequency zero.
(The second-order scalar light shift is identical for both clock levels and does not
contribute to the clock shift).

Consider, for example, a transition between  hfs components 
which have different total angular momenta $F$ and  the
 same projection $M_F$.
 As seen from (\ref{shift}), the frequency shift for such a transition
 can be turned to zero practically
for any frequency of laser field by controlling the orientation
of the external magnetic field. This ``magic'' direction is given by
\begin{equation}
  \cos{\theta}(\omega) = - \frac{2}{A M_F} \frac{\alpha^s_{F_2}(\omega)-
\alpha^s_{F_1}(\omega)}{\alpha^a_{F_2}(\omega)/F_2-\alpha^a_{F_1}(\omega)/F_1}
\label{cos}
\end{equation}
Here we neglect the tensor term which is small for cesium.
%However, it is not neglected in actual calculations.
The numerator of (\ref{cos}) is due
to the third-order (second in the laser field and first in the hfs)
scalar polarizabilities. Second-order polarizabilities (without the hfs) do
not contribute to the energy difference because they do not depend on $F$.
By contrast, the denominator
is strongly dominated by the second-order vector polarizabilities.
The formulas for second and third-order polarizabilities can be found below.

\begin{figure}[!hbt]
\centering
\epsfig{figure=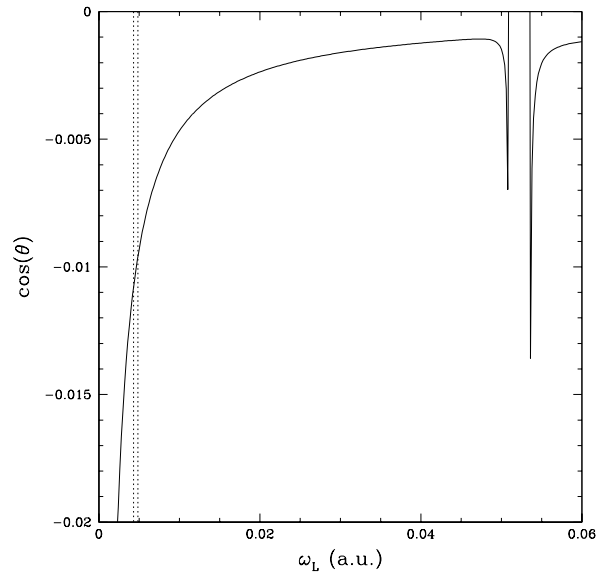,scale=0.4}
\caption{Calculated $\cos(\theta)$ for $^{133}$Cs,
where $\theta$ is the
angle between magnetic field and the light propagation which makes the laser
frequency to be {\em magic}. Vertical lines correspond to frequencies of the CW CO$_2$ laser. }
\label{Fig:Cs}
\end{figure}

Because of the extra hfs operator in the numerator of (\ref{cos}) $\cos(\theta)$
is small which means that magnetic field is directed almost perpendicular to
the propagation of the laser light. Fig.~\ref{Fig:Cs} shows the results of
calculations for the clock transition in cesium between hfs states of
$F_1=3$ and $F_2=4$ with $M_F=3$ in both cases. The results for $\cos(\theta)$
show complicated behavior in the vicinity of the $6s-np_{1/2}$ and
$6s-np_{3/2}$ resonances. First two $6s-6p_{1/2}$ and $6s-6p_{3/2}$
resonances are included
in Fig~\ref{Fig:Cs}. However, everywhere far from resonances $\cos(\theta)$ is
a smooth function of the frequency of laser field. The angle $\theta$ which
makes the frequency to be {\em magic} is always close to $90^o$. 
As $\omega \rightarrow 0$, $\cos{\theta}(\omega) \propto 1/\omega$, and
the angle changes
rapidly at small frequencies which may be advantageous. Indeed, the rapid
change of $\theta$ means that the magic frequency is less sensitive
to uncertainties in $\theta$. Two close dotted vertical lines on Fig.~\ref{Fig:Cs}
show frequencies of the CO$_2$ laser ($\omega_L = 0.43$ a.u. and
$\omega_L = 0.48$ a.u.) which fell into this region. Corresponding values of
$\theta$ are $\theta=90.62^o$ and $\theta=90.55^o$.

\begin{figure}[!hbt]
\centering
\epsfig{figure=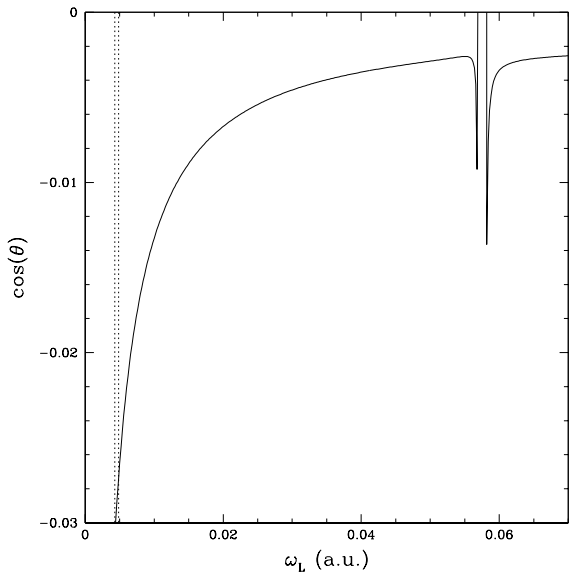,scale=0.4}
\caption{Calculated $\cos(\theta)$ for $^{87}$Rb.}
% where $\theta$ is the
% angle between magnetic field and light propagation which makes the laser
% frequency to be {\em magic}.}
\label{Fig:Rb}
\end{figure}

Fig.~\ref{Fig:Rb} shows the results of similar calculations for the $F=1, M_F=1$
to $F=2, M_F=1$ hfs transition in the ground $5s_{1/2}$ state of $^{87}$Rb.
All notations are the same as for Fig.~\ref{Fig:Cs}. {\em Magic} angles for
the frequencies of the CO$_2$ laser are $\theta=91.77^o$ and $\theta=91.57^o$.

As usual, the linear Zeeman shift may be eliminated if we average the
 frequencies of the hyperfine
transitions with the opposite sign of $M_F$, for example $M_F=3$ and
 $M_F=-3$. However, to keep the cancellation of light shifts in place,
we must have the same value of the products $A M_F \cos{\theta}$
in both transitions. This may be achieved by simultaneous change
of signs of  $M_F$ and circular polarization $A$.

Another solution for eliminating the linear Zeeman shift 
is employing the hyperfine components with $M_{F_1}=-M_{F_2}$. The equation for the magic
angle would be similar to Eq.~(\ref{cos}), except the difference of the vector
polarizabilities in the denominator would be replaced by their sum.

A more accurate treatment of the effects of the B-field requires simultaneous consideration of the magnetic
 interaction and vector light shift. The latter is equivalent to a
 ``pseudo-magnetic'' field directed along the light wave vector $\bf{k}$.
The effective Hamiltonian may be presented in the following form:
\begin{equation}
H=-\frac{{\bf F \cdot Q}}{F},
\end{equation}
where
%\begin{equation}
\[ {\bf Q}=\mu_F {\bf B} + V {\bf n}_k, \]
%\end{equation}
and
%\begin{equation}
\[ V=\left(  \frac{1}{2}\mathcal{E}\right)^{2}A~\alpha_{nF}^{a}\left(
\omega\right)  \frac{1}{2} . \]
%\end{equation}
Here ${\bf n}_k={\bf k}/k$ is the unit vector along ${\bf k}$ and
$\mu_F$ is the magnetic moment of the hyperfine component $F$. The energy
 levels of this Hamiltonian are given by
\begin{equation}\label{E}
E=-\frac{M_F Q}{F},
\end{equation}
where
\begin{eqnarray} \nonumber
 Q=\sqrt{ (\mu_F  B)^2 + 2 V \mu_F  B\cos{\theta} +V^2} \approx \nonumber \\
\mu_F  B +  V \cos{\theta}+V^2/(2 \mu_F  B) + ... \nonumber
\end{eqnarray}
A new feature here is the quadratic vector term which should be subtracted
from the result of measurements to find an accurate value of the
 hyperfine transition frequency.

 Note that in an optical lattice one actually has a standing wave. In this case
it is appropriate to talk about the direction of the photon spin
 ${\bf S}_{ph}=A{\bf n}_k$
 (or the direction of rotation of the light electric field)
 instead of the  direction of the wave vector ${\bf n}_k={\bf k}/k$.
 Indeed, to have the needed standing wave, in the reflected wave
 both ${\bf n}_k$ and $A$ change  sign
  while ${\bf S}_{ph}=A{\bf n}_k$ and the direction of rotation of the light
 electric field are the same.

%{\em Calculation of polarizabilities ---}
\paragraph{Calculation of polarizabilities.}
The second-order dynamic vector polarizability is given by
\begin{eqnarray*}
\lefteqn{\alpha_{n\left(  IJ\right)  F}^{a}\left(  \omega\right)  =  
-\sqrt{\frac{24F\left(  2F+1\right)  }{\left(  F+1\right)  }}  ~\left\{
\begin{array}
[c]{ccc}%
J & F & I\\
F & J & 1
\end{array}
\right\}  ~\times~}\\
& \omega\sum_{n^{\prime}J^{\prime}}(-1)^{J^{\prime}+I+F}\left\{
\begin{array}
[c]{ccc}%
1 & 1 & 1\\
J & J & J^{\prime}%
\end{array}
\right\}  \frac{\left\vert \langle nJ||D||n^{\prime}J^{\prime}\rangle
\right\vert ^{2}}{\left(  E_{nJ}-E_{n^{^{\prime}}J^{^{\prime}}}\right)
^{2}-\omega^{2}} \, .
\end{eqnarray*}

The third-order AC Stark shift (involving two light fields and hfs) is given by the 
frequency-dependent generalization of the DC Stark shift presented previously in 
Refs.~\cite{AngDzuFla06,AngDzuFla06a,BelSafDer06} in the context of the black-body 
radiation clock shift. The structure of the resulting expressions is presented in 
Ref.~\cite{BelDerDzu08}.
Explicit formulas are lengthy  and will be presented elsewhere.

To perform the calculations we follow the procedure described in detail in our
previous work~\cite{AngDzuFla06a}. We use {\em ab initio} relativistic Hartree-Fock 
method in the frozen-core approximation to construct an effective single-electron 
Hamiltonian $\hat H_0$.
Then an all-order correlation potential $\hat \Sigma^{(\infty)}$~\cite{DzuFlaSus89b}
is used to build a complete set of single-electron basis states. 
These states are the eigenstates of the $\hat H_0 + \hat \Sigma^{(\infty)}$ Hamiltonian and
are usually referred to as the Brueckner orbitals. Due to the $\hat \Sigma^{(\infty)}$
operator they include the dominant polarization interaction between core and valence 
electrons.
This approach gives fraction of a per cent accuracy for the energies of valence
states. To calculate matrix elements of the electric dipole and hyperfine
interactions we also include the effect of core polarization by external
field. This is done by means of the time-dependent Hartree-Fock method~\cite{DzuFlaSus84},
which is equivalent to the well-known random-phase approximation.
The resulting accuracy for the polarizabilities is about 1\% (see Ref.~\cite{AngDzuFla06a}
for detailed discussion).

%{\em Conclusion ---}
\paragraph*{Conclusion.}
We conclude that microwave clocks (Cs, Rb,...) using specially engineered ``magic'' 
optical lattices may be
an interesting alternative to the fountain clocks. We have shown
that,  in principle,  the light shift  and the linear Zeeman shift may be
eliminated.  The atoms in the lattice are confined to a tiny volume. This
may help in solving such problems as a homogeneity of magnetic field and cooling 
the chamber to reduce the  thermal black-body radiation shift.
However, the expected accuracy of the optical-lattice microwave clocks is yet to be 
explored. This problem deserves further theoretical and experimental investigation. 
Even if the precision of such clocks turns out be less competitive
than that of the fountains, the microwave lattice clocks have a clear
advantage of a smaller apparatus size. This may  be important for many
applications, e.g., for the spacecraft  applications including navigation
systems and precision tests of fundamental symmetries in space.

%\section*{Acknowledgments}

%{\em Acknowledgments---} 
\paragraph*{Acknowledgments.} 
This work was supported in part by the Australian Research Council,
by the US National Science Foundation and by the US
National Aeronautics and Space Administration under
Grant/Cooperative Agreement No. NNX07AT65A issued by the Nevada NASA
EPSCoR program.

%\bibliography{all}

\end{document}